\documentclass[12pt,preprint]{aastex}

\shorttitle{A new outburst in LMC-N66}
\shortauthors{Pe\~na et al.}

\begin{document}


\title{A New Outburst In the Extraordinary Central Star of LMC-N66\altaffilmark{1,2,3}}


\author{Miriam Pe\~na\altaffilmark{4}, M. Teresa Ruiz\altaffilmark{5}, Patricio Rojo\altaffilmark{5} }

\author{Silvia Torres-Peimbert\altaffilmark{4} and Wolf-Rainer Hamann\altaffilmark{6}}


\altaffiltext{1}{ESO proposals ID 279.D-5024 and ID 076.B-0166}
\altaffiltext{2}{Based on observations made with the NASA-CNES-CSA Far Ultraviolet Spectroscopic Explorer (FUSE), operated for NASA by the John Hopkins University under NASA contract NAS5-32985.}
\altaffiltext{3}{Based on observations obtained at Las Campanas (Carnegie).}
\altaffiltext{4}{Instituto de Astronom{\'\i}a, Universidad Nacional Aut\'onoma de M\'exico, Apdo. Postal 70264, M\'exico, D.F., 04510, M\'exico. E-mail: miriam@astroscu.unam.mx}
\altaffiltext{5}{Departamento de Astronom{\'\i}a, Universidad de Chile, Casilla 36D, Santiago, Chile}
\altaffiltext{6}{Universit\"at Potsdam, Am Neuen Palais 10, 14469 Potsdam, Germany}


\begin{abstract}
This is the first report on the new outburst presented by the central star of the LMC-N66 nebula. This object was classified as a planetary nebula, however, its true nature is under debate. In the period 1955$-$1990 the central star was  almost undetectable and only nebular emission lines were observed. In 1990, the beginning of an outburst was detected and in few months it became much brighter and developed wide He and N lines, typical of  a Wolf Rayet star  of the N-sequence. The maximum occurred in 1994 and afterwards the star slowly faded. Analysis of its evolution showed that it has a variable mass-loss rate which occasionally increases enormously, creating a false photosphere at a much larger radius, making it appear a few magnitudes brighter.  The present outburst has occurred 13 years after the  episode from 1994 to 2000. So far this new event has similar characteristics although there are some significant differences in the spectral features. We present optical and FUSE spectra showing the main properties of this latter event.
\end{abstract}



\keywords{stars: winds, outflows -- stars: mass-loss -- stars: Wolf-Rayet -- planetary nebulae: individual: LMC-N66}


\section{Introduction}

In 1993, a huge mass-loss event was detected in the central star of the planetary nebula LMC-N66 (also  SMP 83 and WS 35). The star developed features of a Wolf-Rayet star (WR) of the N-sequence and its brightness increased by a large factor while the nebular characteristics remained unchanged (Pe\~na et al. 1994; Pe\~na et al. 1995). The event was monitored at UV and optical wavelengths. The mass-loss  gradually decreased and the star returned to its quiet state in about 6 years. A detailed analysis of the stellar evolution, using advanced non-LTE models for expanding atmospheres fitting the UV and optical data across these epochs, was performed by Hamann et al. (2003), who showed that the stellar luminosity increased from log ($L/L_\odot$)\,=\,4.6 at the end of 1990 (quiescence) to 5.4 in 1994 and then returned gradually to the initial value. In the same period the mass-loss rate increased (quiescence) 10$^{-5.7}$ M$_\odot$ yr$^{-1}$  to 10$^{-5.0}$ M$_\odot$ yr$^{-1}$ at maximum, while the stellar temperature remained constant during the whole event (T$_*$ = 112 $\pm$ 20 kK). The chemical composition  of the ejecta corresponded to that of an incompletely CNO-processed material: dominated by helium (X$_{\rm He}$= 0.8) with a small amount of hydrogen (X$_{\rm H}$=0.2), nitrogen slightly enhanced (X$_{\rm N}$=3 E-03) and carbon very depleted (X${\rm _C} \leq$ 1 E-04). Several possible scenarios for the nature  of LMC-N66 (single massive star, post-AGB star, high-mass binary system with mass transfer, or white dwarf accreting mass) are discussed by Hamann et al. Although none of the scenarios are completely satisfactory, it is suggested that the most probable ones are those involving an evolved binary system. In particular, LMC-N66 could be a white dwarf accreting matter at a high rate in a close binary. The large  accretion may bring the stellar mass to the Chandrasekhar limit within a few hundred thousand years, making it a candidate for a Type Ia supernova precursor.

Due to its extraordinary characteristics we have kept monitoring this  object at irregular intervals over the years. The last spectrum  obtained that showed no outburst was taken  in January 28, 2006. Fifteen months later,  on April 26, 2007, a new outburst was discovered.

\section{Observations}
 Once the outburst was detected, a follow-up program was started right away, thanks to a Director Discretionary Time (DDT) program, with EMMI and the NTT at European Southern Observatory (ESO). Concurrently we obtained DDT with the FUSE satellite. 
In this section we describe these observations.

\subsection{Optical Spectrum}

A spectrum obtained with the ESO VLT and FORS1 spectrograph in January 28, 2006 shows the star  in quiescence, presenting  a faint continuum and  faint \ion{He}{2} 4686 wide component.  The outburst was discovered in April 26, 2007 with an optical spectrum of N66 obtained at Las Campanas using the duPont telescope equipped with the B\&C spectrograph.
  Right after the discovery, ESO DDT was granted through program ID 279.D-5024. The ESO-NTT + EMMI spectrograph with a combination of gratings has been used to gather spectroscopic data in the 3250$-$8000 \AA\AA \ wavelength range. Spectra were obtained on July 22, August 14, August 30, September 14, and October 22, 2007. The exposure times were of 20 min in the blue and 10 min in the red. 
  
   The spectrum in August 30,  2007 is shown in Fig. 1 together with the data obtained on January 28, 2006 (quiescence) and a spectrum of August 2, 1994 (maximum of previous outburst). Some stellar characteristics, derived from these spectra, are presented in Table~1. It is evident that during outbursts the star    increases its brightness and develops wide and prominent WN features (mainly  \ion{He}{2},  \ion{N}{5} and  \ion{N}{4} lines). For the two observed outbursts the spectra are very similar presenting the same WN features; although the stellar magnitude in 2007, as measured from the continuum flux at 5450 \AA, is brighter than in 1994 for 0.5 mag and about  2.2 mag brighter than in quiescence.  Notice that due to the differences in slit widths, the flux listed for 1994 and 2006 correspond to the whole stellar flux, while in the  2007 observation some of the emission could have been lost due to seeing conditions. The mass-loss rate listed in Table~1 for 1994 was taken from Hamann et al. (2003) and  for 2006 we adopted the quiescent value derived by Hamann et al.; alternatively, for 2007 we estimated $\dot M$ by scaling the value of 1994, taking into account that F(\ion{He}{2} 4686) is 1.7 times more intense in 2007  and that $\dot M \propto \sqrt F$.

 In 2007 the stellar continuum is  brighter and also the WR emission lines are more intense, which could be understood in terms of a higher mass-loss rate. On the other hand, the WR features present  very similar FWHMs in both outbursts, indicating similar wind velocities. There are other small differences between outbursts, for instance the \ion{N}{5} 3484 \AA \  is stronger in 2007 than in 1994, and  the \ion{C}{4} 5808 doublet is more prominent, relative to \ion{He}{2} 5411 \AA \ and \ion{He}{1}  5876 \AA, as can be observed in Fig. 2 where we present the  5300$-$ 6000 \AA\AA \ zone normalized to the continuum.   It is known that the stellar spectrum of LMC-N66 undergoes small variations in short time scales (of about a month), therefore it seems likely that the observed differences in the spectra between 2007 and 1994 could be due to short time  variations of the wind parameters and  do not necessarily reflect differences in chemical composition or other important parameters.
 
\begin{center}
\begin{table}
\caption{Stellar characteristics for different epochs.\label{tbl-1}}
\begin{tabular}{lrrr}
\tableline\tableline
Obs-date& 02/08/94  & 28/01/06 & 30/08/07 \\ 
\tableline
Slit width ($''$) & 10 & 5 & 1.02\\
F(He II 4686)$^a$ & 4.0E-13 &  4.9E-14: & 6.6E-13 \\
FWHM(He II 4686) (\AA)& 50  &  54: & 49 \\
Stellar cont.$^{b}$& 1.1E-15  &   2.3E-16 & 1.7E-15 \\
M$_{\rm V}$ &    16.3   &  18.0    & 15.8   \\
log $\dot M$ & -5.0  &   -5.7 & -4.9  \\
\tableline
\multicolumn{4}{l}{$^a$ Flux in erg cm$^{-2}$ s$^{-1}$.}\\
\multicolumn{4}{l}{$^b$ Stellar continuum at 5450 \AA, in erg cm$^{-2}$ s$^{-1}$ \AA $^{-1}$. }\\
\multicolumn{4}{l}{\hskip 0.2cm The nebular contribution has been subtracted.} 
\end{tabular}
\end{table}
\end{center}

\subsection{FUV spectrum}
 
Due to the high stellar temperature, the maximum emission from the star  is at short wavelengths, therefore it was very important to secure UV information. FUSE DDT was granted through the program ID Z018. The observations covered the wavelength range 905$-$1187 \AA\AA, at a spectral resolution of R $\sim$ 20,000. Originally an exposure time of 27000 s was approved, but unfortunately the termination of the FUSE science mission on August 2007 made it impossible to complete the full exposure. Only two observations (ID Z0180101000 of 8235 s and Z0180101000 of 5783 s) were acquired on July 2007. Nevertheless the stellar brightness was sufficient to obtain data with significant signal-to-noise. The FUSE LWRS ($30'' \times 30''$) aperture was used. The data, taken in ``tag time'' mode, were calibrated using the FUSE data reduction pipeline.

A combination of the signal obtained through  the four channels (LiF1, LiF2, SiC1, SiC2) of the two detectors (A \& B) is presented in Fig. 3. This spectrum was smoothed to reduce the spectral resolution to 0.2 \AA. Wide emission lines from the central star are clearly appreciated. The most prominent lines correspond to \ion{He}{2}, \ion{S}{6}, \ion{O}{6}, and \ion{P}{5}. According to Willis et al. (2004) these features correspond to a high excitation WN (see  their figures 2a and 2b and their \S 4.2). The most important stellar lines and their characteristics are listed in Table 2. Previous FUSE data  for LMC-N66 central star were published by  Herald \& Bianchi (2007). In comparison with their  spectrum, obtained  in 2003 at quiescent state (see their  Fig. 1), our 2007 spectrum shows a continuum 5  times more intense and well developed WR lines.

In addition to the wide stellar lines, narrow emission lines from the Lyman series are conspicuous. These lines correspond to the airglow, but some nebular lines like \ion{C}{3} 977 \AA, \ion{O}{6} 1032, 1038 \AA\AA \ and \ion{He}{2} 1086 \AA \ are observed. Narrow absorptions displaced 1 \AA \ to the red from the Lyman airglow lines are produced by  interstellar absorption in the LMC (1 \AA \ corresponds to the radial velocity of 300 km s$^{-1}$ measured for LMC-N66 by Pe\~na et al. 2004).

\section{Discussion}
By comparing the characteristics of the far UV stellar lines with those of the FUSE spectra for massive WR stars published by Willis et al. (2004), we find that the spectral classification for the central star of LMC-N66 corresponds to an early WN, probably a WN\,3. In the previous outburst, and based on optical spectra only, we classified it as a WN\,4.5. Certainly this object is not a classical WR star as it is much less luminous, it presents outbursts difficult to understand, and in addition, its spectrum shows short-time variability that yields a slightly different spectral classification.
For 2007, the optical spectrum shows \ion{N}{4} and \ion{He}{1} lines, so the WN\,4.5 classification is still valid. But, as said, the spectral classification for this object is not as direct as in the case of massive WRs  due to its peculiarities.

Undoubtedly the central star of LMC-N66 is undergoing a new huge mass-loss event only 13 years after the previous episode.  In this event, the star appears about 0.5 mag brighter than in the previous one and shows more intense WR features. A period as short as 13 yr for a new outburst  was not expected. Although the star was reported by Nail \& Shapley (1955) as a variable (HV\,5967) with $\Delta$m$\sim$0.9 mag, subsequent variations were not reported for more than 30 years, although this object was observed at least in 1975, 1976, 1983, 1985, 1988, 1989 and 1990 by different authors (e.g., Dopita et al. 1985; Monk et al. 1988; Pe\~na \& Ruiz 1988; Meatheringham \& Dopita 1991; Dopita et al. 1993). None of these observations were made in a year corresponding to a probable maximum in an apparent 13-yr period (going back in time, the years for a maximum would have been 1981, 1968, 1955). But taking into account that the 1994 outburst declined slowly and that the star was a couple of magnitudes brighter than in quiescence and
 the stellar lines were  prominent  for  a few  years after maximum, such features would have been detected had they been present in other epochs.  Therefore there is no evidence for an outburst of the intensity of the ones in 1994 and 2007 to have taken place  from 1955 to 1990.

Thus  there is no evidence for periodicity in  the stellar variations. The LMC-N66 central star can only be classified as an irregular variable with a  mass-loss that can increase enormously occasionally and we cannot predict the next outburst. We  continue to  monitor the star in the optical range, to follow its evolution and  to verify if the spectral differences between 2004 and 1994 persist (see \S 2.1 and Table~1). The data obtained  will be used to compute new atmosphere models to detect any variation in the stellar parameters.

\acknowledgments

M. Pe\~na is grateful to DAS, Universidad de Chile, for hospitality during a sabbatical stay when part of this work was performed. This work received financial support from DGAPA-UNAM (IN-118405) and CONACyT (46904). M.T. Ruiz and P. Rojo acknowledge support from the FONDAP Center for Astrophysics and from the Centro de Astrof{\'\i}sica y Tecnolog{\'\i}as Asociadas (CATA).



\begin{center}
\begin{table}
\caption{Far UV stellar WR emission features.\label{tbl-2}}
\begin{tabular}{lrccl}
\tableline\tableline
$\lambda_0$ (\AA) & ion &  Flux$^a$ & comments \\ 
\tableline
933.4 & \ion{S}{6} &  6.54 &  P-Cygni\\
944.5 & \ion{S}{6}  &  6.46 & P-Cygni \\
958.7 & \ion{He}{2}  &   3.41 & blended with \ion{N}{4}? \\
 977.0 & \ion{C}{3} &  5.29 &  nebular \ion{C}{3}  \\
 992.3 & \ion{He}{2}  &   3.72 & nebular \ion{He}{2}  \\
 1031.8+37.5 &\ion{O}{6}   &   4.38 & nebular \ion{O}{6}  \\
 1085.7 & \ion{He}{2} & 3.74 &  nebular \ion{He}{2}   \\
 1118.0 & \ion{P}{5}  &  0.50 &  P-Cygni \\
 1128.0 & \ion{P}{5}  & 1.10&  P-Cygni \\
\tableline
\multicolumn{4}{l}{$^a$ In units of 10$^{-13}$ erg cm$^{-2}$ s$^{-1}$ \AA $^{-1}$.}\\
\end{tabular}
\end{table}
\end{center}

\clearpage

\begin{figure*}
\plotone{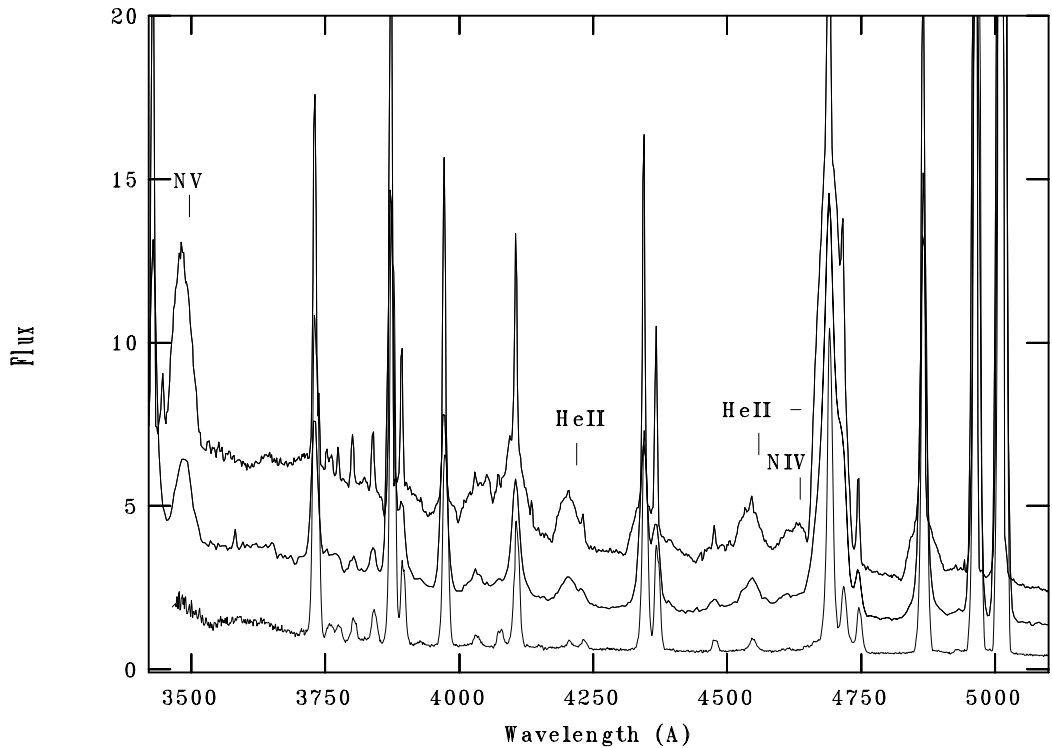}
\caption{Optical spectra of LMC-N66 for 30 August 2007 (top), 1994 (middle) and 2006 (bottom) are presented, showing the new outburst, the previous one and the quiescent state.  Flux is in units of 10$^{-15}$ erg cm$^{-2}$ s$^{-1}$ \AA $^{-1}$.  The most the most intense WR features are marked. The 2007 spectrum shows  a brighter stellar continuum and  more intense WR features probably due to a larger mass-loss rate.   \label{fig1}}
\end{figure*}

\begin{figure}
\epsscale{.80}
\plotone{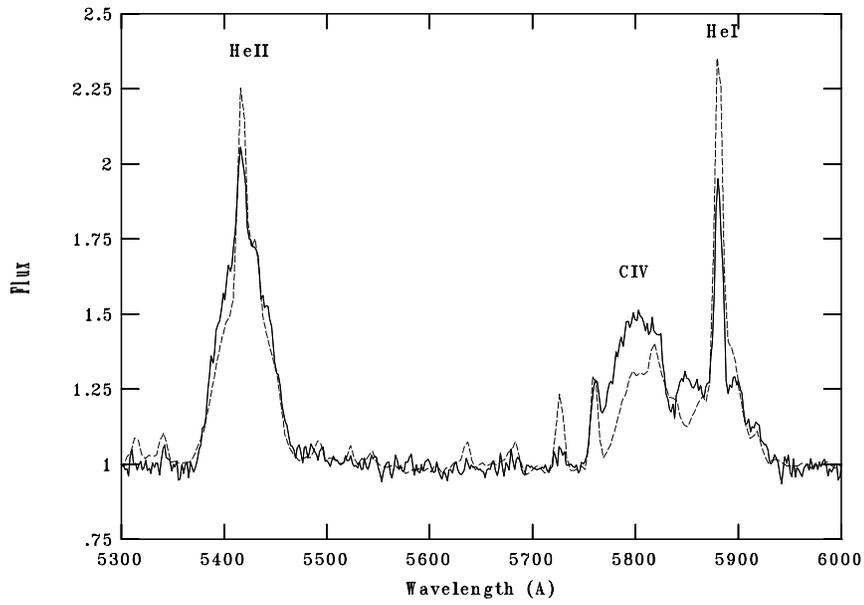}
\caption{Comparison of  \ion{He}{2}  5411, \ion{He}{1} 5876, and \ion{C}{4} 5808 features for 2007 (solid line) and 1994 (dashed line) observations. The spectra have been normalized to compare the wide emission features.The \ion{C}{4} 5808 line is much brighter, relative to the He lines, in the 2007 event.\label{fig2}}
\end{figure}

\begin{figure}
\epsscale{1.0}
\plotone{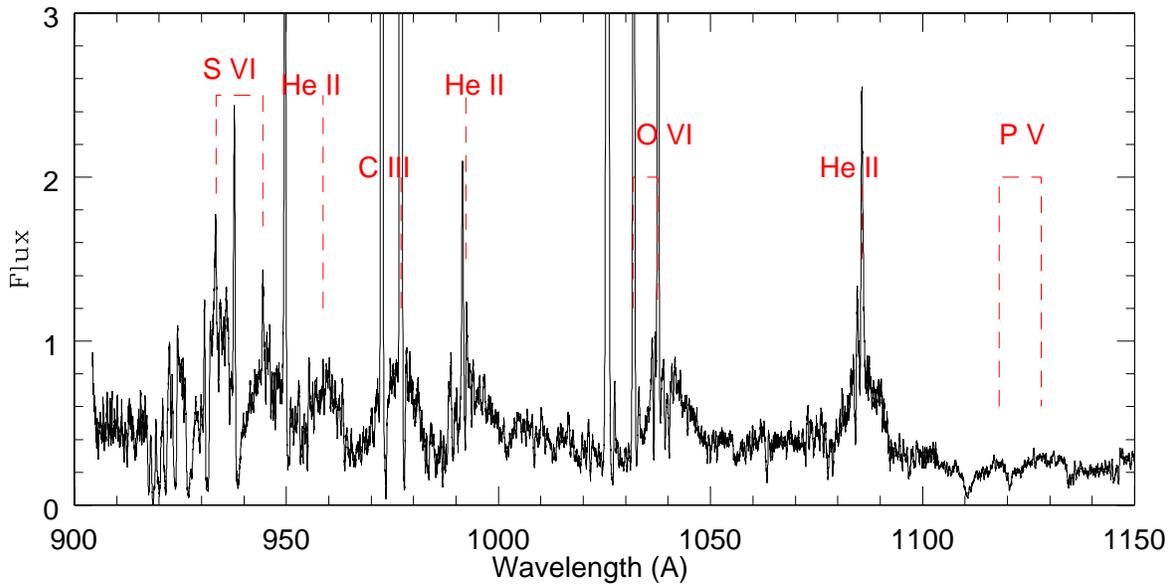}
\caption{FUSE spectrum  of the central star of LMC-N66, in July 2007. Flux in units of 10$^{-13}$ erg cm$^{-2}$ s$^{-1}$ \AA $^{-1}$. The marked  WR  features are similar to the ones of a WN\,3 star (see Willis et al.  2004). Narrow  nebular lines of the Lyman series (produced by airglow) and nebular  \ion{C}{3} (977 \AA),  \ion{He}{2} (992 \AA), \ion{O}{6} (1032,1038 \AA\AA) and \ion{He}{2} (1086 \AA) are also very intense.\label{fig3}} 
\end{figure}

\end{document}